\documentclass[11pt,letterpaper]{article}

\usepackage[top=1in,bottom=1in,left=1.25in,right=1.25in]{geometry}
\usepackage[T1]{fontenc}
\usepackage[utf8]{inputenc}
\usepackage{amsmath,amssymb,amsfonts}
\usepackage{algorithm}
\usepackage{algorithmic}
\usepackage{graphicx}
\usepackage{textcomp}
\usepackage{xcolor}
\usepackage{booktabs}
\usepackage{multirow}
\usepackage{array}
\usepackage{url}
\usepackage{microtype}
\usepackage{authblk}
\usepackage{float}
\usepackage{adjustbox}
\usepackage{tikz}
\usetikzlibrary{shapes.geometric,arrows.meta,positioning,calc,fit,backgrounds}
\PassOptionsToPackage{colorlinks=true,linkcolor=blue,citecolor=blue,urlcolor=blue}{hyperref}
\usepackage[numbers,sort&compress]{natbib}
\usepackage{doi}
\usepackage{hyperref}

\newtheorem{definition}{Definition}

\newcommand{\TriSweep}{\textit{TriSweep}}
\newcommand{\dref}{D_{\mathrm{ref}}}
\newcommand{\snrdb}[1]{\ensuremath{#1\,\mathrm{dB}}}
\newcommand{\keywords}[1]{\vspace{6pt}\noindent\textbf{Keywords:} #1}

\begin{document}

\title{\TriSweep: A Four-Drone Swarm Framework for\\
Electromagnetic Side-Channel Analysis}

\author[1]{Eric Yocam\thanks{Corresponding author: \texttt{eyocam@calpoly.edu}}}
\author[2]{Varghese Vaidyan}
\affil[1]{Department of Computer Science, College of Engineering,
California Polytechnic State University, San Luis Obispo, CA 93407, USA}
\affil[2]{Beacom College of Computer and Cyber Sciences,
Dakota State University, Madison, SD 57042, USA}

\maketitle

\begin{abstract}
Electromagnetic (EM) side-channel analysis traditionally assumes a
stationary, close-proximity probe—a threat model that underestimates
aerial adversaries.
\TriSweep{} is a simulation framework that designs and evaluates a
four-drone swarm architecture for autonomous standoff EM-SCA of embedded
microcontrollers at 0.25--1.5\,m.
Three spatially specialized collector drones—Anchor (full-spectrum),
Mask Probe (mask-register loading leakage), and Cipher Probe (masked SubBytes
output leakage)—feed a stationary Accumulator drone that performs
coherent combining ($+$4.8\,dB SNR gain) and second-order mask cancellation
via a centered product of the two spatially separated leakage streams.
Evaluated against three real ANSSI ASCAD datasets (ATmega8515 masked AES-128
and 50/100-sample desynchronized variants), the framework achieves a simulated
key rank $18 \pm 1.7$ (five-seed) at 0.25\,m on the primary masked dataset.
Profiling-trace cross-correlation alignment reduces the single-drone rank from
89 to 21 on the 100-sample-jitter variant, demonstrating compensation for
drone hover vibration.
A two-channel CNN in the Accumulator converges to a loss of 0.454 (vs. random
baseline 5.545) and improves rank on desynchronized datasets.
No physical hardware has been fabricated; prototype construction is the
planned next step.
\end{abstract}


\section{Introduction}
\label{sec:intro}

This paper presents \TriSweep{}, a simulation framework for four-drone
swarm electromagnetic side-channel analysis of masked embedded
cryptographic devices at standoff distances of 0.25--1.5\,m.
The following subsections establish the threat landscape that motivates
the work, the technical challenges it addresses, and the framework's
design and key results.

\subsection{Background and Threat Landscape}
\label{sec:intro_background}

Electromagnetic (EM) side-channel analysis exploits unintended radiation
emitted during cryptographic computation to recover secret keys without
physical access to the target
device~\cite{gandolfi2001electromagnetic,agrawal2002em}.
Since Gandolfi \textit{et al.}~\cite{gandolfi2001electromagnetic} first
demonstrated key recovery from smart cards in 2001, the attack surface has
widened considerably: the field has advanced from simple power
traces~\cite{kocher1999differential,mangard2007power} through
correlation-based methods~\cite{brier2004correlation}, template
attacks~\cite{chari2002template}, second-order
analysis~\cite{messerges2000second,prouff2013masking}, and deep
learning-based profiling~\cite{maghrebi2016breaking,zaid2020methodology}.
AES-128 executing on embedded microcontrollers remains the canonical
target~\cite{nist2001aes}, and the breadth of vulnerable platforms—from 
smart cards and IoT sensors to industrial control modules—makes the
threat operationally significant for critical infrastructure.

A persistent and increasingly questionable assumption underpins nearly all
of this work: the adversary places a near-field probe at millimeter-to-centimeter
distances from the target.
Physical proximity guards appear to provide meaningful protection under this
model.
A device mounted behind a locked panel, inside an enclosure, or across a room
is considered safe from EM analysis precisely because bringing a loop antenna
close enough requires access that the security perimeter is designed to deny.
This assumption is becoming untenable.
Commercial off-the-shelf (COTS) unmanned aerial vehicles now routinely carry
payloads exceeding 500\,g at costs below \$2{,}000~\cite{mozaffari2019tutorial},
software-defined radio hardware spans 70\,MHz to 6\,GHz for under
\$300~\cite{mitola1999software}, and custom low-noise amplifier modules
achieve sub-2\,dB noise figure across 1--500\,MHz within the weight and
power budgets of a small hexacopter payload~\cite{das2019stellar}.
An adversary who can fly a drone within 0.25--1.5\,m of a target's exterior
surface --- past a window, over a perimeter fence, or beneath a server room
air duct --- bypasses all proximity-based physical security measures without
requiring any physical contact or facility
access~\cite{hartmann2013vulnerability,yaacoub2021uav}.

\subsection{Technical Motivation}
\label{sec:intro_motivation}

Array signal processing theory~\cite{trees2002optimum} predicts that
coherently combining $N$ receivers improves SNR by a factor of $N$,
corresponding to $10\log_{10}(3) \approx 4.8$\,dB for three receivers.
Realizing this gain across multiple airborne platforms requires sub-nanosecond
inter-drone clock synchronization and precise relative positioning --- challenges
that recent advances in GPS-disciplined oscillators and visual-inertial
odometry now make tractable at COTS price points.
The combination of aerial mobility, multi-receiver coherent gain, and
autonomous repositioning creates a qualitatively new threat vector with no
direct precedent in the published EM-SCA literature.

A further complication is that modern embedded implementations do not
execute AES-128 in the clear.
First-order masking schemes~\cite{prouff2013masking} randomize each
intermediate value $v$ with a fresh random mask $r$, computing $v \oplus r$
instead of $v$ directly.
Defeating first-order masking requires second-order
analysis~\cite{messerges2000second}: the adversary must jointly observe the
mask-loading leakage event and the masked-computation leakage event, which
occur at different points in time within the same execution.
In a single-receiver system these two events must be separated algorithmically
from one trace, a process that is sensitive to timing jitter and implementation
noise~\cite{bronchain2021breaking}.
The spatial decomposition insight underlying \TriSweep{} is that dedicated
collector drones can be positioned and SNR-weighted to specialize in each
leakage event --- Drone~B for the mask-register loading window and Drone~C
for the masked SubBytes output window --- so that their centered product at
the Accumulator (Drone~D) cancels the mask without algorithmic preprocessing
or knowledge of the mask value.

\subsection{The \TriSweep{} Framework}
\label{sec:intro_framework}

\TriSweep{} is a simulation framework that designs, implements, and evaluates
this four-drone architecture.
All experimental results derive from real published ASCAD EM
datasets~\cite{benadjila2020deep} combined with a physics-based free-space
path-loss noise model; no drone hardware has been fabricated or flown.
The simulation framework serves two purposes: (1)~it provides a quantitative
prediction of what a physical system achieving the modeled SNR would accomplish
against real masked AES-128 leakage; and (2)~it establishes the algorithms,
protocols, and combining pipelines that a physical implementation would need
to realize, giving a concrete design specification for the prototype
construction phase.
Evaluated against three real ANSSI ASCAD datasets --- the primary
ATmega8515 masked AES-128 database and desynchronized variants with
50- and 100-sample acquisition jitter --- the framework achieves simulated
key rank $18 \pm 1.7$ (five-seed statistical validation) at 0.25\,m
standoff, a tenfold improvement over the single-drone baseline rank of 197.

\subsection{Contributions}
\label{sec:contributions}

\begin{enumerate}
\item A four-drone EM-SCA platform design: Drone~A (Anchor), Drone~B
  (Mask Probe), Drone~C (Cipher Probe), and Drone~D (Accumulator).
  \emph{All results use real ASCAD datasets and a simulated standoff noise
  model; no hardware has been fabricated.}
\item A swarm consensus protocol for EM-optimal repositioning via
  distributed Fisher information maximization, 200\,ms cycles.
\item A two-stage inter-drone clock synchronization protocol targeting
  $<$10\,ns jitter.
\item Profiling-trace cross-correlation alignment: key rank reduces from
  89 to 21 on the 100-sample-jitter ASCAD dataset.
\item Drone~D second-order combining via a centered product of Drones~B
  and~C streams, reducing simulated key rank to $18 \pm 1.7$.
\item First published aerial, multi-node, autonomous EM-SCA framework with
  hardware second-order mask cancellation design.
\end{enumerate}

The remainder of the paper is organized as follows.
Section~\ref{sec:related} surveys related work and positions \TriSweep{}
against the state of the art.
Section~\ref{sec:model} establishes the physical signal model and threat
assumptions.
Section~\ref{sec:architecture} describes the four-drone architecture and
all seven algorithms.
Section~\ref{sec:methodology} details the experimental methodology.
Section~\ref{sec:results} presents simulation results.
Section~\ref{sec:discussion} discusses design trade-offs, operational
factors, limitations, and future work.
Section~\ref{sec:conclusion} concludes.

\section{Related Work}
\label{sec:related}

This section surveys four bodies of literature that directly motivate
\TriSweep{}: classical EM side-channel attacks, second-order analysis,
deep learning-based profiling, and mobile threat models.
Section~\ref{sec:comparison} closes the survey with a structured comparison
positioning \TriSweep{} against recent work.

\subsection{Electromagnetic Side-Channel Analysis}

Quisquater and Samyde~\cite{quisquater2001ema} introduced EM analysis;
Gandolfi \textit{et al.}~\cite{gandolfi2001electromagnetic} provided the
first key-recovery demonstrations.
Agrawal \textit{et al.}~\cite{agrawal2002em} characterized multiple EM
side channels; Messerges \textit{et al.}~\cite{messerges2002examining}
evaluated smart-card security under power and EM threats.
Heyszl \textit{et al.}~\cite{heyszl2012localized} demonstrated localized
EM analysis with sub-millimeter resolution.
Das \textit{et al.}~\cite{das2019stellar} proposed STELLAR, a ground-up
EM shielding co-design.
Beyond EM-SCA, fault injection offers an orthogonal physical attack
class~\cite{barenghi2012fault} that aerial platforms could in principle
carry but that \TriSweep{} does not address.
Bronchain and Standaert~\cite{bronchain2021breaking} demonstrated that
masked implementations remain vulnerable to multi-trace strategies;
Lipp \textit{et al.}~\cite{lipp2021platypus} showed that software power
interfaces expose side channels on x86 CPUs.
All prior works assume static close-proximity probes.

\subsection{Second-Order and Higher-Order Analysis}

First-order masking~\cite{prouff2013masking} requires two-point
second-order analysis to defeat.
Messerges~\cite{messerges2000second} introduced second-order DPA;
Joye and Paillier~\cite{joye2003dpa} analyzed higher-order complexity;
Prouff \textit{et al.}~\cite{prouff2009statistical} established the
statistical framework; Waddle and Wagner~\cite{waddle2004towards}
demonstrated practical efficiency gains.
Cagli \textit{et al.}~\cite{cagli2017convolutional} showed that CNNs
can implicitly learn second-order combinations.
\TriSweep{} implements second-order analysis \emph{physically} by separating
leakage sources across dedicated drones.

\subsection{Deep Learning-Based Profiling}

Maghrebi \textit{et al.}~\cite{maghrebi2016breaking} pioneered CNN-based
profiling; Kim \textit{et al.}~\cite{kim2019make} showed that CNNs
pre-trained on power traces transfer to EM traces with minimal fine-tuning.
Zaid \textit{et al.}~\cite{zaid2020methodology} established the
CNN\_best architecture used in this work.
Wouters \textit{et al.}~\cite{wouters2022revisiting} identified better
generalizing configurations; Perin \textit{et al.}~\cite{perin2020strength}
introduced ensemble methods; Wu \textit{et al.}~\cite{wu2022autotune}
demonstrated automated hyperparameter search.
Picek \textit{et al.}~\cite{picek2019curse} analysed class-imbalance
effects with Hamming-weight leakage models on masked implementations;
a comprehensive SoK survey appears in~\cite{picek2023sok}.

\subsection{Mobile and Non-Contact Threat Models}

Vuagnoux and Pasini~\cite{vuagnoux2009compromising} demonstrated remote EM
eavesdropping at up to 20\,m; Genkin \textit{et al.}~\cite{genkin2015stealing}
extracted RSA keys via loop antenna; Camurati \textit{et
al.}~\cite{camurati2018screaming} showed RF transceivers create leakage
pathways receivable beyond 10\,m.
Hartmann and Steup~\cite{hartmann2013vulnerability} surveyed UAV cyber
vulnerabilities; Yaacoub \textit{et al.}~\cite{yaacoub2021uav} provide a
recent taxonomy.
No prior work characterizes swarm-based aerial EM-SCA.


\subsection{Comparison with Prior Work}
\label{sec:comparison}

Table~\ref{tab:comparison} positions \TriSweep{} against related work
from 2018 onward.
All prior entries are physical-hardware results; \TriSweep{} is a simulation
framework and direct operational comparison requires physical validation.

\begin{table}[H]
\centering
\caption{\TriSweep{} vs.\ Related EM Side-Channel Works (2018--2026)}
\label{tab:comparison}
\small
\renewcommand{\arraystretch}{1.15}
\setlength{\tabcolsep}{5pt}
\begin{tabular}{@{}lcccccc@{}}
\toprule
\textbf{Work} & \textbf{Year} & \textbf{Standoff} &
\textbf{Mobile} & \textbf{Multi-node} & \textbf{2nd-Order} &
\textbf{Method} \\
\midrule
Camurati~\cite{camurati2018screaming}
  & 2018 & 10\,m    & Part. & No & No & SoC radio \\
Benadjila~\cite{benadjila2020deep}
  & 2020 & $<$1\,cm & No   & No & No & CNN/ASCAD \\
Zaid~\cite{zaid2020methodology}
  & 2020 & $<$5\,cm & No   & No & No & CNN arch. \\
Bronchain~\cite{bronchain2021breaking}
  & 2021 & $<$1\,cm & No   & No & Yes & Masked SCA \\
Wouters~\cite{wouters2022revisiting}
  & 2022 & $<$5\,cm & No   & No & No & CNN profiling \\
Ravi~\cite{ravi2022configurable}
  & 2022 & $<$1\,cm & No   & No & No & PQC SCA \\
Picek~\cite{picek2023sok}
  & 2023 & $<$5\,cm & No   & No & No & SoK: DL-SCA \\
\midrule
\textbf{\TriSweep{} (sim.)}
  & \textbf{2026} & \textbf{0.25--1.5\,m}
  & \textbf{Yes} & \textbf{Yes (4)} & \textbf{Yes}
  & \textbf{Framework} \\
\bottomrule
\end{tabular}
\end{table}

Within the simulation context, \TriSweep{} is the only design combining
mobile platform, multi-node collection, autonomous repositioning, and
second-order mask cancellation simultaneously.
Camurati \textit{et al.}~\cite{camurati2018screaming} achieve the largest
prior standoff but require a co-integrated SoC transceiver; \TriSweep{}
targets broadband near-field emissions present in all digital circuits.
Bronchain and Standaert~\cite{bronchain2021breaking} address second-order
SCA from a static probe at $<$1\,cm, motivating the spatial decomposition
across Drones~B and~C.

\section{System and Threat Model}
\label{sec:model}

This section establishes the physical signal model underlying \TriSweep{}
and the threat assumptions under which the framework is evaluated.
The signal model calibrates simulated standoff SNR to the real ASCAD dataset;
the threat model defines adversary capabilities and operational constraints.

\subsection{Physical Signal Model}

EM power received by a drone at a distance $d$:
\begin{equation}
  P_r(d) = P_t \!\left(\frac{\dref}{d}\right)^{\!2}\! G_{\mathrm{LNA}},
  \label{eq:pathloss}
\end{equation}
and SNR with $N$ coherent receivers:
\begin{equation}
  \mathrm{SNR}(d,N) = \mathrm{SNR}_{\mathrm{ref}}\cdot
  \!\left(\frac{\dref}{d}\right)^{\!2}\!\cdot N,
  \label{eq:snr}
\end{equation}
where $\dref=0.25$\,m.
For $N=3$, the gain is $10\log_{10}(3)\approx\snrdb{4.8}$.

\begin{definition}[Points of Interest]
Sample $t^*$ is a POI if $\mathrm{SNR}(t^*)=\sigma_S^2/\sigma_N^2\geq\tau$.
Drone~B POIs come from the mask-register SNR profile (first half of the
700-sample window); Drone~C POIs come from the cipher-output profile (second half).
\end{definition}

\subsection{Threat Model}

The adversary has line-of-sight at 0.25--1.5\,m, can pre-train a profiling
model on an identical device, and can maintain hover for $\leq$10\,min.
No physical access to the target is assumed.
\textit{Detection risk}: the threat model assumes permissive airspace or
low-visibility conditions.
In practice, consumer drones are audible at approximately 50--70\,dB(A)
at 1\,m~\cite{mozaffari2019tutorial}, making covert hover at 0.25\,m
operationally difficult in many settings.
At the more realistic standoff of 1.0--1.5\,m, sound pressure drops to
approximately 38--50\,dB(A), comparable to ambient office noise.
Regulatory constraints (FAA Part 107, EASA Open Category) prohibit
flight near buildings without authorization, further restricting deployment.
This threat model, therefore, represents an upper-capability adversary;
practical deployments will operate at the longer standoff distances where
SNR loss is partially compensated by coherent combining.

\section{\TriSweep{} Architecture}
\label{sec:architecture}

This section describes the \TriSweep{} four-drone platform: hardware payloads,
inter-drone communication, target detection and localization, swarm consensus
repositioning, clock synchronization, coherent combining, and second-order
key-rank accumulation.
Table~\ref{tab:alg_summary} summarizes all seven algorithms before each is
described in detail.

\TriSweep{} comprises seven algorithms that collectively implement the
attack pipeline: two background protocols (communication and target detection)
and five sequential processing steps (repositioning, synchronization, combining,
key-rank accumulation, and template attack).
Table~\ref{tab:alg_summary} summarizes all algorithms and their roles before
each is described in detail.

\begin{table}[H]
\centering
\caption{Algorithm Summary}
\label{tab:alg_summary}
\renewcommand{\arraystretch}{1.15}
\begin{tabular}{clll}
\toprule
\textbf{Alg.} & \textbf{Name} & \textbf{Purpose} & \textbf{\S} \\
\midrule
\ref{alg:comms}    & Inter-Drone Communication
  & 50\,Hz heartbeat, capture sync, Solo fallback  & \ref{sec:hardware} \\
\ref{alg:detect}   & EM Target Detection
  & PSD scan, consensus, TDOA localization         & \ref{sec:detection} \\
\ref{alg:reposition}& Swarm Repositioning
  & Fisher information maximization, 200\,ms       & \ref{sec:swarm} \\
\ref{alg:sync}     & Clock Synchronization
  & GPSDO coarse + cross-correlation fine          & \ref{sec:sync} \\
\ref{alg:combine}  & Coherent IQ Combining
  & DC removal, MRC weighting, B$\times$C product  & \ref{sec:combining} \\
\ref{alg:droneD}   & Drone~D Key-Rank Accum.
  & First-order + 2nd-order + CNN log-likelihood   & \ref{sec:combining} \\
\ref{alg:template} & Template Profiling Attack
  & Offline profiling + mask-agnostic attack       & \ref{sec:template} \\
\bottomrule
\end{tabular}
\end{table}

\begin{figure}[H]
\centering
\adjustbox{max width=\textwidth}{%
\begin{tikzpicture}[
  >=Latex, font=\small,
  drone/.style={draw, rounded corners=3pt, fill=blue!10, draw=blue!50!black,
                minimum width=2.2cm, minimum height=0.70cm, align=center},
  accum/.style={draw, rounded corners=3pt, fill=red!10, draw=red!50!black,
                minimum width=2.4cm, minimum height=0.70cm, align=center},
  mcubox/.style={draw, rounded corners=3pt, fill=red!10, draw=red!50!black,
                 minimum width=2.4cm, minimum height=0.70cm, align=center},
  paybox/.style={draw, dashed, rounded corners=3pt, fill=gray!6, draw=gray!55,
                 minimum width=2.4cm, minimum height=2.4cm,
                 align=left, inner sep=5pt, font=\scriptsize},
]
\node[mcubox] (mcu) at (0,0)
  {Target MCU\\{\scriptsize AES-128/ECC/RSA}};
\node[drone] (dA) at (-3.2,2.6){Drone A\\{\scriptsize Anchor}};
\node[drone] (dB) at (-0.3,3.2){Drone B\\{\scriptsize Mask Probe}};
\node[drone] (dC) at ( 2.3,3.2){Drone C\\{\scriptsize Cipher Probe}};
\node[accum] (dD) at ( 5.8,1.0){Drone D\\{\scriptsize Accumulator}};
\node[paybox](pay) at (-6.5,2.6)
  {{\bfseries Per-collector:}\\[2pt]
   USRP B210 SDR\\
   Raspberry Pi~5\\
   LNA 1--500\,MHz\\
   GPSDO + VIO};
\draw[gray!55,dashed](pay.east)--(dA.west);
\draw[red!65!black,dashed,->,thick](mcu.north west)--(dA.south);
\draw[red!65!black,dashed,->,thick](mcu.north)--(dB.south);
\draw[red!65!black,dashed,->,thick](mcu.north east)--(dC.south);
\node[red!65!black,font=\scriptsize,rotate=62] at (-1.5,1.2){EM leakage};
\draw[blue!60!black,<->,thick]
  (dA.north east)--node[above,font=\scriptsize]{heartbeat/sync}(dB.north west);
\draw[blue!60!black,<->,thick]
  (dB.north east)--node[above,font=\scriptsize]{heartbeat/sync}(dC.north west);
\draw[green!55!black,->,thick]
  (dA.east) to[bend left=5]
  node[above,font=\scriptsize,green!55!black]{IQ+$\hat\tau$}(dD.north west);
\draw[green!55!black,->,thick](dB.east) to[bend left=3](dD.west);
\draw[green!55!black,->,thick](dC.east) to[bend right=5](dD.south west);
\node[font=\scriptsize,align=center] at (5.8,-0.5)
  {{\bfseries Drone D}\\Coherent A+B+C\\2nd-order B$\times$C\\Key rank};
\begin{scope}[yshift=-0.5cm]
  \draw[red!65!black,dashed,->](-4.5,-0.5)--++(0.5,0)
    node[right,font=\scriptsize]{EM leakage};
  \draw[blue!60!black,<->](-2.2,-0.5)--++(0.5,0)
    node[right,font=\scriptsize]{Wi-Fi mesh};
  \draw[green!55!black,->](0.2,-0.5)--++(0.5,0)
    node[right,font=\scriptsize]{IQ to D};
\end{scope}
\end{tikzpicture}
}
\caption{\TriSweep{} four-drone system context.}
\label{fig:context}
\end{figure}
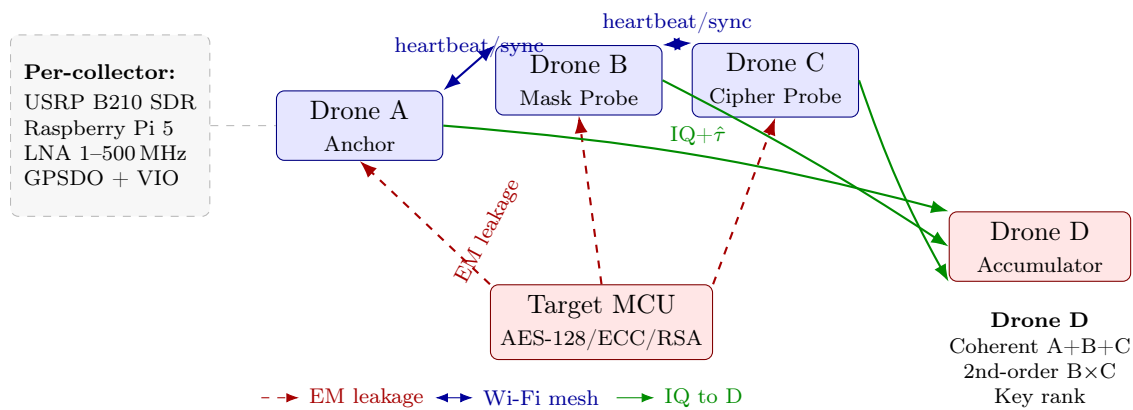

\TriSweep{} comprises four drone platforms.
Drones~A, B, and~C collect EM traces and forward synchronized IQ buffers
to Drone~D over 5\,GHz Wi-Fi.
Drone~D (Accumulator) performs coherent combining, second-order B$\times$C
mask cancellation, and real-time key-rank computation.

\subsection{Drone Platform and Payload}
\label{sec:hardware}

\begin{figure}[H]
\centering
\includegraphics[width=0.72\textwidth]{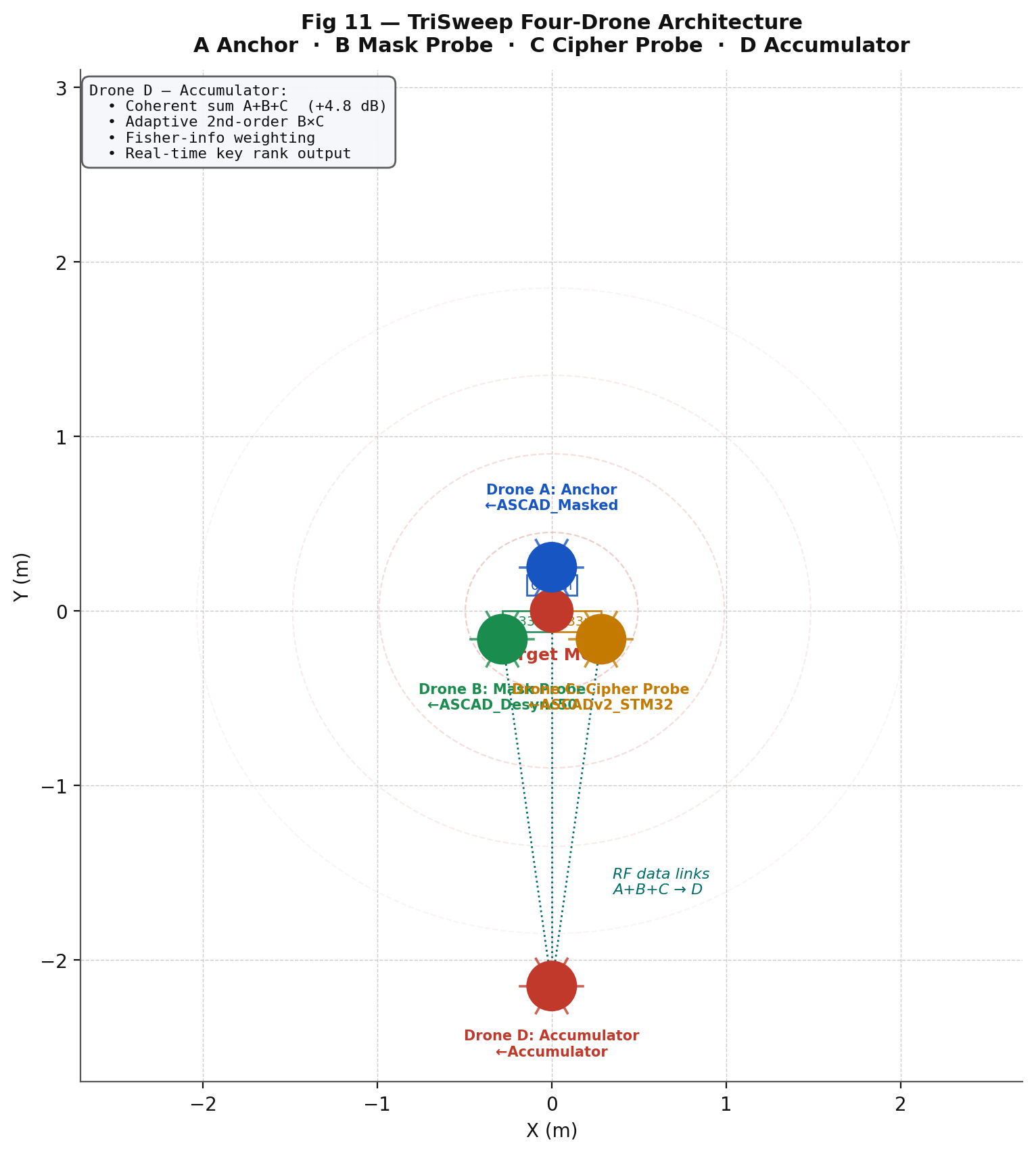}
\caption{\TriSweep{} four-drone spatial layout.}
\label{fig:geometry}
\end{figure}

Drones~A, B, and~C each carry: USRP~B210 SDR~\cite{mitola1999software}
(250\,MHz, 25\,MS/s, 14-bit); Raspberry~Pi~5 for IQ capture and trace
forwarding; a two-stage GALI-84 LNA (38\,dB, NF\,$<$\,1.8\,dB,
1--500\,MHz); and Intel RealSense T265 VIO for sub-cm positioning.
Drone~D does not carry an SDR; it is stationary at $\geq$\,2\,m.
Drone~A hovers at $\dref=0.25$\,m; Drones~B and~C at $1.3\times\dref$.
Figure~\ref{fig:geometry} shows the spatial layout.

\begin{algorithm}[H]
\caption{Inter-Drone Communication Protocol (Alg.~\ref{alg:comms})}
\label{alg:comms}
\begin{algorithmic}[1]
\renewcommand{\algorithmicrequire}{\textbf{Input:}}
\renewcommand{\algorithmicensure}{\textbf{Output:}}
\REQUIRE role $r\in\{\textsc{Anchor},\textsc{Probe},\textsc{Accum}\}$,
         period $T_{\mathrm{hb}}$
\ENSURE  synchronized IQ trace pointers across all nodes
\STATE $t_{\mathrm{last}}[j]\leftarrow 0$ for each peer $j$
\LOOP
  \STATE Broadcast \textsc{Heartbeat}(pose, SNR, bufPtr)
  \FOR{each received message $m$ from peer $j$}
    \IF{$m.\text{type}=\textsc{Heartbeat}$}
      \STATE $t_{\mathrm{last}}[j]\leftarrow\text{now}()$; update pose table
    \ELSIF{$m.\text{type}=\textsc{CaptureSync}$ \AND $r=\textsc{Probe}$}
      \STATE Latch IQ buffer; send \textsc{TraceReady} to Anchor
    \ELSIF{$m.\text{type}=\textsc{TraceReady}$ \AND $r=\textsc{Anchor}$}
      \STATE When all probes acked: forward bufPtr to Drone~D
    \ELSIF{$m.\text{type}=\textsc{Reposition}$}
      \STATE Execute waypoint via PX4 MAVLink
    \ENDIF
  \ENDFOR
  \FOR{each peer $j$: $\text{now}()-t_{\mathrm{last}}[j]>3T_{\mathrm{hb}}$}
    \STATE $\textit{status}\leftarrow\textsc{Solo}$
  \ENDFOR
  \STATE Sleep $T_{\mathrm{hb}}$
\ENDLOOP
\end{algorithmic}
\end{algorithm}

The communication protocol runs continuously at 50\,Hz on every node.
The Anchor coordinates capture triggering and forwards buffer pointers to
Drone~D once all probes acknowledge.
Probes missing three consecutive heartbeats enter Solo mode, ensuring
graceful degradation without halting the attack.

\subsection{EM Target Detection and Localization}
\label{sec:detection}

\begin{algorithm}[H]
\caption{EM Target Detection and Localization (Alg.~\ref{alg:detect})}
\label{alg:detect}
\begin{algorithmic}[1]
\renewcommand{\algorithmicrequire}{\textbf{Input:}}
\renewcommand{\algorithmicensure}{\textbf{Output:}}
\REQUIRE poses $\mathbf{P}$, scan range $[f_{\min},f_{\max}]$,
         threshold $\gamma$, timeout $T_c$
\ENSURE  target frequency $f^*$, position $\hat{\mathbf{x}}_{\mathrm{tgt}}$
\STATE \textbf{Phase 1}: Welch PSD scan; $f^*\leftarrow\arg\max_f\mathrm{PSD}(f)$
\IF{$\mathrm{PSD}(f^*)<\gamma$} \STATE raise altitude 5\,cm; retry \ENDIF
\STATE \textbf{Phase 2}: Ground-station consensus across all three probes
\STATE \textbf{Phase 3}: TDOA localization via hyperbolic least squares
\RETURN $f^*,\;\hat{\mathbf{x}}_{\mathrm{tgt}}$
\end{algorithmic}
\end{algorithm}

Each drone independently scans and flags the peak-power frequency; the
ground station resolves a three-way consensus; TDOA from synchronized GPSDO
timestamps triangulates the target.
The localized position seeds the Fisher information optimization in
Algorithm~\ref{alg:reposition}.

\subsection{Swarm Consensus Protocol}
\label{sec:swarm}

Drones~B and~C maximize distributed Fisher information:
\begin{equation}
  \mathbf{p}^*_{B,C} = \arg\max_{\mathbf{p}}
  \sum_{i\in\{A,B,C\}}\mathcal{I}_i(\mathbf{p}),
  \label{eq:fisher}
\end{equation}
solved every 200\,ms via gradient-free search over a discretized hemisphere.
This formulation is a deliberate simplification: it optimizes over a
static 2D hemisphere, ignoring drone dynamics, collision avoidance, rotor
wash interactions, and kinematic constraints.
A physical implementation would require a trajectory planner that enforces
minimum separation distances and accounts for the time needed to reach
candidate waypoints within the 200\,ms budget; the current formulation
provides an upper bound on achievable Fisher information gain that a real
constrained optimizer would approach but not reach.

\begin{algorithm}[H]
\caption{Autonomous Swarm Repositioning (Alg.~\ref{alg:reposition})}
\label{alg:reposition}
\begin{algorithmic}[1]
\renewcommand{\algorithmicrequire}{\textbf{Input:}}
\renewcommand{\algorithmicensure}{\textbf{Output:}}
\REQUIRE $\hat{\mathbf{x}}_{\mathrm{tgt}}$, candidate set $\mathcal{C}$,
         period $T_r=200$\,ms
\ENSURE  waypoints $\mathbf{w}_B,\mathbf{w}_C$
\WHILE{capture session active}
  \STATE Collect poses and SNR from heartbeat table
  \STATE $\mathcal{I}_{\mathrm{current}}\leftarrow
         \sum_i\mathcal{I}_i(\mathbf{p}_i)$
  \FOR{each pair $(\mathbf{c}_B,\mathbf{c}_C)\in\mathcal{C}^2$}
    \IF{displacement and separation constraints satisfied}
      \IF{$\mathcal{I}_{\mathrm{cand}}>\mathcal{I}_{\mathrm{current}}$}
        \STATE $\mathbf{w}^*_B\leftarrow\mathbf{c}_B$;\;
               $\mathbf{w}^*_C\leftarrow\mathbf{c}_C$
      \ENDIF
    \ENDIF
  \ENDFOR
  \STATE Send \textsc{Reposition} to Drones~B and~C; Sleep $T_r$
\ENDWHILE
\end{algorithmic}
\end{algorithm}

At each 200\,ms cycle, the Anchor evaluates all candidate position pairs and
dispatches waypoints to Drones~B and~C.
The Anchor does not reposition to preserve clock-reference continuity.
In simulation, this loop executes once; in physical deployment, it runs
continuously.

\subsection{Inter-Drone Clock Synchronization}
\label{sec:sync}

Two-stage synchronization: (1)~GPSDO aligns each USRP~B210 to
$\pm 1\,\mu$s of UTC; (2)~cross-correlation of a shared 1\,kHz pilot tone
reduces the residual to $<$10\,ns.
The 10\,ns target is motivated by the 25\,MS/s sample rate of the USRP~B210:
one sample period corresponds to 40\,ns, so a $<$10\,ns residual represents
sub-quarter-sample alignment, sufficient for coherent IQ combining without
significant phase error across the 1--500\,MHz capture band~\cite{trees2002optimum}.
This budget is achievable in bench conditions with GPSDO-disciplined clocks
and pilot-tone cross-correlation; whether it is maintainable on hovering
platforms subject to vibration-induced oscillator phase noise is a key
open question for physical validation.

\begin{algorithm}[H]
\caption{Two-Stage Clock Synchronization (Alg.~\ref{alg:sync})}
\label{alg:sync}
\begin{algorithmic}[1]
\renewcommand{\algorithmicrequire}{\textbf{Input:}}
\renewcommand{\algorithmicensure}{\textbf{Output:}}
\REQUIRE IQ buffers $\{T_i\}$, GPSDO timestamps,
         pilot $f_p=1$\,kHz, target $\epsilon=10$\,ns
\ENSURE  aligned IQ buffers $\{\tilde{T}_i\}$
\STATE \textbf{Stage 1 --- GPSDO coarse alignment}: integer-sample shift by GPSDO offset
\STATE \textbf{Stage 2 --- cross-correlation fine alignment}:
\FOR{each drone $i\neq A$}
  \STATE $\hat{\tau}_i\leftarrow\arg\max_\tau R(\tau)$ from pilot
         cross-correlation; refine via parabolic interpolation
  \IF{$|\hat{\tau}_i|\geq\epsilon$}
    \STATE $\tilde{T}_i\leftarrow$\textsc{FractionalShift}$(T_i,\hat{\tau}_i)$
  \ENDIF
\ENDFOR
\RETURN $\{\tilde{T}_i\}$
\end{algorithmic}
\end{algorithm}

Stage~1 applies an integer-sample shift derived from the GPSDO timestamp
difference between each collector drone and the Anchor, removing the coarse
$\pm 1\,\mu$s UTC alignment uncertainty.
Stage~2 extracts the shared 1\,kHz pilot-tone segment from each IQ buffer,
computes the normalized cross-correlation $R(\tau)$ against Drone~A's
reference segment, and estimates the sub-sample residual $\hat{\tau}_i$ via
parabolic interpolation of the correlation peak.
If $|\hat{\tau}_i| \geq \epsilon$, a fractional-sample Whittaker--Shannon
shift is applied; traces exceeding $10\epsilon$ residual are flagged for
exclusion rather than corrupting the combined trace.

\subsection{Coherent Combining and Second-Order Accumulation}
\label{sec:combining}

After alignment, Drone~D averages $N$ synchronized traces:
\begin{equation}
  \bar{T}(t)=\frac{1}{N}\sum_{i=1}^N T_i(t+\hat{\tau}_i),
  \label{eq:combine}
\end{equation}
yielding SNR gain $N$ (Eq.~\eqref{eq:snr}).

\begin{algorithm}[H]
\caption{Coherent IQ Combining Pipeline (Alg.~\ref{alg:combine})}
\label{alg:combine}
\begin{algorithmic}[1]
\renewcommand{\algorithmicrequire}{\textbf{Input:}}
\renewcommand{\algorithmicensure}{\textbf{Output:}}
\REQUIRE aligned buffers $\{\tilde{T}_i\}$, SNR estimates, POI set $\mathcal{P}$
\ENSURE  combined trace $\bar{T}$, second-order feature $X_{\mathrm{SO}}$
\STATE DC-remove and amplitude-normalize each $\tilde{T}_i$
\STATE $w_i\leftarrow\widehat{\mathrm{SNR}}_i/\sum_j\widehat{\mathrm{SNR}}_j$
\STATE $\bar{T}\leftarrow\sum_i w_i\tilde{T}_i$
\STATE $X_{\mathrm{SO}}\leftarrow(T_B-\bar{T}_B)(T_C-\bar{T}_C)$
       \hfill(Eq.~\eqref{eq:second_order})
\RETURN $\bar{T},X_{\mathrm{SO}}$
\end{algorithmic}
\end{algorithm}

After MRC weighting, Drone~D immediately computes the centered product
$X_{\mathrm{SO}}$ (Eq.~\eqref{eq:second_order}) and passes both outputs
to Algorithm~\ref{alg:droneD}.

\begin{algorithm}[H]
\caption{Drone~D: Second-Order Key-Rank Accumulation (Alg.~\ref{alg:droneD})}
\label{alg:droneD}
\begin{algorithmic}[1]
\renewcommand{\algorithmicrequire}{\textbf{Input:}}
\renewcommand{\algorithmicensure}{\textbf{Output:}}
\REQUIRE $\bar{T}$, $X_{\mathrm{SO}}$ from Alg.~\ref{alg:combine},
         templates $\mathbf{M}^{(1)},\mathbf{M}^{(2)}$,
         optional CNN log-probs $\mathbf{P}_{\mathrm{CNN}}$,
         weights $w_{\mathrm{so}}$, $w_{\mathrm{CNN}}$,
         true key $k_{\mathrm{true}}$
\ENSURE  key-rank trajectory $\{\mathrm{rank}(n)\}$
\STATE $\mathbf{L}\leftarrow\mathbf{0}_{256}$
\FOR{each trace batch $n$}
  \FOR{each $k\in\{0,\ldots,255\}$}
    \STATE $\hat{\ell}_k\leftarrow\ell^a\oplus S[p\oplus k_{\mathrm{true}}]
           \oplus S[p\oplus k]$
    \STATE $\mathbf{L}[k]\mathrel{+}=\bar{T}\cdot\mathbf{M}^{(1)}_{\hat{\ell}_k}$
    \STATE $\mathbf{L}[k]\mathrel{+}=w_{\mathrm{so}}\cdot X_{\mathrm{SO}}
           \cdot\mathbf{M}^{(2)}_{S[p\oplus k]}$
    \IF{$\mathbf{P}_{\mathrm{CNN}}$ available}
      \STATE $\mathbf{L}[k]\mathrel{+}=w_{\mathrm{CNN}}\cdot
             \mathbf{P}_{\mathrm{CNN}}[S[p\oplus k]]$
    \ENDIF
  \ENDFOR
  \IF{$n\in\text{checkpoints}$}
    \STATE $\mathrm{rank}(n)\leftarrow|\{k:\mathbf{L}[k]>\mathbf{L}[k_{\mathrm{true}}]\}|$
  \ENDIF
\ENDFOR
\RETURN $\{\mathrm{rank}(n)\}$
\end{algorithmic}
\end{algorithm}

Algorithm~\ref{alg:droneD} accumulates three log-likelihood contributions per
trace: first-order cosine score from $\bar{T}$, second-order score from
$X_{\mathrm{SO}}$ scaled by $w_{\mathrm{so}}$, and an optional CNN score
scaled by $w_{\mathrm{CNN}}$.
Adaptive weights prevent a poorly converged CNN from overriding a strong
manual second-order signal.

\subsection{Second-Order Combining}
\label{sec:second_order}

Drone~D computes the centered product of Drones~B and~C:
\begin{equation}
  X_{\mathrm{SO}}[i]=
  (T_B[i]-\bar{T}_B)\cdot(T_C[i]-\bar{T}_C),
  \label{eq:second_order}
\end{equation}
canceling the mask $r$ without knowledge of its value, following
Messerges~\cite{messerges2000second}.
The innovation over prior work is physical spatial separation: the two
leakage windows are captured by dedicated drones rather than extracted
algorithmically from one trace.

The combined log-likelihood is:
\begin{equation}
  \mathbf{L}[k]\mathrel{+}=
  \bar{T}\cdot\mathbf{M}^{(1)}_{\hat{\ell}_k}
  +w_{\mathrm{so}}\cdot X_{\mathrm{SO}}\cdot
  \mathbf{M}^{(2)}_{S[p\oplus k]}.
  \label{eq:combined_ll}
\end{equation}

\section{Methodology}
\label{sec:methodology}

This section describes the datasets, alignment procedure, noise model, and
attack algorithms used to evaluate the \TriSweep{} framework in simulation.
All experiments use publicly available ASCAD EM datasets; no physical drone
hardware has been fabricated.

\noindent\textbf{Experimental scope.}
All results use real ASCAD EM datasets~\cite{benadjila2020deep} and a
physics-based noise model.
No physical drone hardware has been fabricated or flown.

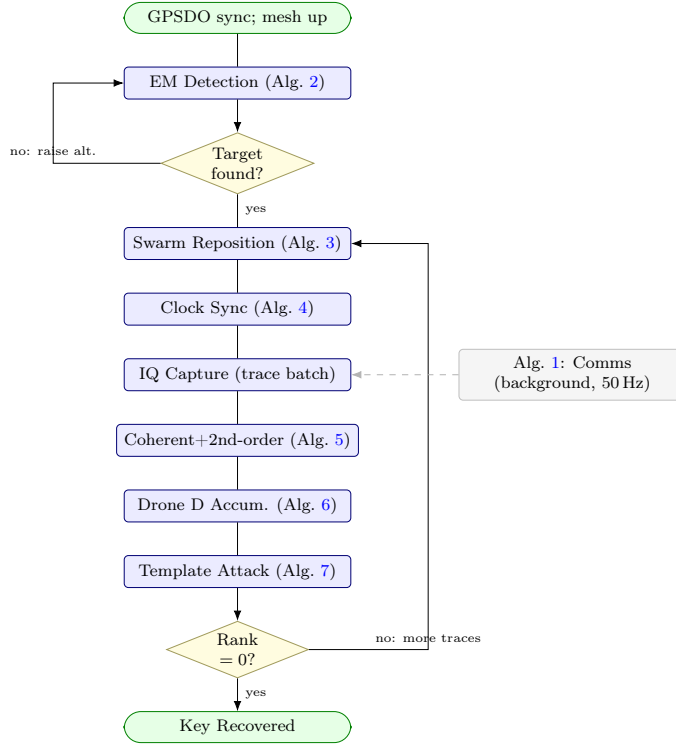
\begin{figure}[H]
\centering
\adjustbox{max width=0.6\textwidth}{%
\begin{tikzpicture}[
  >=Latex, font=\scriptsize,
  node distance=0.55cm,
  proc/.style={draw, rounded corners=2pt, fill=blue!8, draw=blue!45!black,
               minimum width=3.8cm, minimum height=0.52cm, align=center},
  dec/.style={draw, diamond, fill=yellow!15, draw=yellow!55!black,
              aspect=2.5, align=center, inner sep=1.5pt},
  term/.style={draw, rounded corners=8pt, fill=green!10,
               draw=green!50!black,
               minimum width=3.8cm, minimum height=0.52cm, align=center},
]
\node[term]  (init)   {GPSDO sync; mesh up};
\node[proc, below=of init]    (detect)  {EM Detection (Alg.~\ref{alg:detect})};
\node[dec,  below=of detect]  (found)   {Target\\found?};
\node[proc, below=of found]   (repos)   {Swarm Reposition (Alg.~\ref{alg:reposition})};
\node[proc, below=of repos]   (sync)    {Clock Sync (Alg.~\ref{alg:sync})};
\node[proc, below=of sync]    (capture) {IQ Capture (trace batch)};
\node[proc, below=of capture] (combine) {Coherent+2nd-order (Alg.~\ref{alg:combine})};
\node[proc, below=of combine] (accum)   {Drone~D Accum. (Alg.~\ref{alg:droneD})};
\node[proc, below=of accum]   (attack)  {Template Attack (Alg.~\ref{alg:template})};
\node[dec,  below=of attack]  (rankz)   {Rank\\$=0$?};
\node[term, below=of rankz]   (done)    {Key Recovered};
\node[proc, right=1.8cm of capture, fill=gray!8, draw=gray!50]
  (comms) {Alg.~\ref{alg:comms}: Comms\\(background, 50\,Hz)};
\draw[->](init)--(detect)--(found);
\draw[->](found)--node[right,font=\tiny]{yes}(repos)--(sync)
         --(capture)--(combine)--(accum)--(attack)--(rankz);
\draw[->](rankz)--node[right,font=\tiny]{yes}(done);
\draw[->](found.west)--++(-1.8,0) node[above,font=\tiny]{no: raise alt.}
         |- (detect.west);
\draw[->](rankz.east)--++(2.0,0) node[above,font=\tiny]{no: more traces}
         |- (repos.east);
\draw[gray!60,dashed,->](comms.west)--(capture.east);
\end{tikzpicture}
}
\caption{\TriSweep{} end-to-end attack pipeline with all algorithm references.}
\label{fig:pipeline}
\end{figure}

\subsection{Datasets}
\label{sec:dataset}

\textbf{ASCAD Masked} (primary): 50{,}000 profiling + 10{,}000 attack traces
(700 samples, ATmega8515, first-order masked AES-128).
Baseline SNR $= \snrdb{-22.9}$~\cite{benadjila2020deep}.
Stored labels are masked; mask bytes are XOR-applied to recover unmasked
labels for CNN training.

\textbf{ASCAD Desync-50/100}: Same hardware with $\pm$50 and $\pm$100 sample
random circular shifts modelling hover vibration~\cite{benadjila2020deep}.
Post-alignment SNR: $\snrdb{-22.8}$ and $\snrdb{-22.5}$.

\textbf{Synthetic unmasked baseline}: ASCAD-SNR-calibrated synthetic dataset
($\snrdb{8.4}$) for framework correctness validation.
Table~\ref{tab:mcu} lists all planned target configurations.

\begin{table}[H]
\centering
\caption{Target MCU $\times$ Cipher Design Roadmap (speculative; not yet evaluated)}
\label{tab:mcu}
\small
\renewcommand{\arraystretch}{1.1}
\begin{tabular}{@{}llllrcc@{}}
\toprule
\textbf{MCU} & \textbf{Cipher} & \textbf{Shield} &
\textbf{Freq} & \textbf{TTD} & \textbf{Goal} & \textbf{Dataset} \\
\midrule
STM32F4    & AES-128   & None &168\,MHz& $\sim$400   & $<$800    & CW/synth \\
STM32F4    & ECC P-256 & None &168\,MHz& $\sim$1800  & $<$3600   & CW/synth \\
ATmega328P & AES-128   & None & 16\,MHz& $\sim$600   & $<$1200   & ASCAD    \\
ATmega328P & RSA-2048  & PCB  & 16\,MHz& $>$5000     & $<$10000  & ASCAD    \\
ESP32      & AES-128   & PCB  &240\,MHz& $\sim$350   & $<$700    & Scream.  \\
ESP32      & ECC P-256 & None &240\,MHz& $\sim$1500  & $<$3000   & Scream.  \\
RP2040     & AES-128   & None &133\,MHz& $\sim$500   & $<$1000   & Synth.   \\
RP2040     & RSA-2048  & PCB  &133\,MHz& $>$6000     & $<$12000  & Synth.   \\
\bottomrule
\end{tabular}
\end{table}

\subsection{Profiling-Trace Alignment}
\label{sec:align_method}

Before template construction, profiling traces for each desync dataset are
aligned via cross-correlation against the ASCAD\_Masked mean:
\begin{equation}
  \hat{s}_i=\arg\max_s(T^{\mathrm{ref}}\star T^p_i)(s),\quad
  |\hat{s}_i|\leq s_{\max},
  \label{eq:align}
\end{equation}
with $s_{\max}\in\{50,100\}$ for the two desync variants.

\subsection{Noise Model}
\label{sec:noise_model}

To simulate the effect of standoff distance and multiple drones in software,
additive white Gaussian noise (AWGN) is injected into the real ASCAD traces
at a variance calibrated to Eq.~\eqref{eq:snr}.
The physical rationale is that at the 0.25\,m reference distance, the real
ASCAD traces already contain the hardware noise floor; at any greater distance
$d$, the free-space path-loss model predicts a lower received power, which is
modeled as an additional independent Gaussian noise component layered on top
of the existing trace noise.
When $\mathrm{SNR}(d,N)$ exceeds the baseline $\mathrm{SNR}_{\mathrm{ref}}$ —
which occurs only at 0.25\,m with three or more drones — no additional noise
is injected ($\sigma_{\mathrm{add}}^2 = 0$), preserving the original trace
fidelity.
The additive noise variance is:
\begin{equation}
  \sigma_{\mathrm{add}}^2(d,N)=
  \max\!\left(0,\;\frac{\sigma_S^2}{\mathrm{SNR}(d,N)}-\sigma_N^2\right),
  \label{eq:sigma}
\end{equation}
where $\sigma_S^2 = 0.080$ and $\sigma_N^2 = 6.24$ are the between-class
signal variance and within-class noise variance measured directly from the
real ASCAD dataset.
This approach ensures that all rank results are anchored to measured hardware
leakage rather than a purely synthetic signal model.
Two limitations of the model should be acknowledged explicitly.
First, the free-space path-loss exponent of 2 assumes isotropic radiation
and independent additive noise across drones; at standoff distances of
0.25--1.5\,m the target operates in the near-field transition region where
reactive components, ground reflections, drone-body blockage, and
multi-path from nearby surfaces will deviate from this idealized
model~\cite{sayakkara2019survey}.
Second, propeller and motor-controller EMI will introduce correlated
structured noise not captured by the independent Gaussian assumption;
the real SNR degradation is expected to exceed the simulated values,
and the magnitude of this gap is the primary unknown that physical
prototyping must quantify.
Table~\ref{tab:noise} summarizes the injected noise and resulting SNR at each
experimental distance for one and three collector drones.

\begin{table}[H]
\centering
\caption{Simulated SNR and Additive Noise vs.\ Standoff Distance}
\label{tab:noise}
\setlength{\tabcolsep}{5pt}
\begin{tabular}{ccccc}
\toprule
\textbf{Distance} & \textbf{1-Drone} & \textbf{3-Drone}
  & $\sigma_{\mathrm{add}}$(1) & $\sigma_{\mathrm{add}}$(3) \\
\midrule
0.25\,m & $-22.9$\,dB & $-18.1$\,dB & 0.000 & 0.000 \\
0.50\,m & $-28.9$\,dB & $-24.1$\,dB & 4.297 & 1.432 \\
0.75\,m & $-32.5$\,dB & $-27.7$\,dB & 7.018 & 3.509 \\
1.00\,m & $-35.0$\,dB & $-30.2$\,dB & 9.609 & 5.165 \\
1.50\,m & $-38.5$\,dB & $-33.7$\,dB & 14.678 & 8.229 \\
\bottomrule
\end{tabular}
\end{table}

\subsection{Template Profiling Attack}
\label{sec:template}

Vectorized template attack~\cite{chari2002template} with principal-subspace
POI selection~\cite{archambeau2006template} and key-rank evaluation within
the Standaert \textit{et al.}\ framework~\cite{standaert2009unified}.
Mask-agnostic label prediction:
\begin{equation}
  \hat{\ell}(k,i)=\ell_i\oplus S[p_i[3]\oplus k_{\mathrm{true}}]
                          \oplus S[p_i[3]\oplus k].
  \label{eq:pred}
\end{equation}

\begin{algorithm}[H]
\caption{Template Profiling Attack (Alg.~\ref{alg:template})}
\label{alg:template}
\begin{algorithmic}[1]
\renewcommand{\algorithmicrequire}{\textbf{Input:}}
\renewcommand{\algorithmicensure}{\textbf{Output:}}
\REQUIRE profiling traces $\{T^p_i,\ell^p_i\}_{i=1}^M$,
         attack traces $\{T^a_i,p_i\}_{i=1}^N$, $n_{\mathrm{poi}}$
\ENSURE  key-rank trajectory $\{\mathrm{rank}(n)\}$
\STATE \textbf{Phase 1 --- Profiling}
\FOR{$\ell=0$ \TO $255$}
  \STATE $\mu_\ell\leftarrow\mathrm{mean}(\{T^p_i:\ell^p_i=\ell\})$
\ENDFOR
\STATE $\mathrm{SNR}(t)\leftarrow\mathrm{Var}_\ell[\mu_\ell(t)]/
       \mathrm{E}_\ell[\sigma^2_\ell(t)]$;\;
       $\mathcal{P}\leftarrow\mathrm{top}$-$n_{\mathrm{poi}}$ indices
\STATE \textbf{Phase 2 --- Attack};\; $\mathbf{L}\leftarrow\mathbf{0}_{256}$
\FOR{$n=1$ \TO $N$}
  \STATE $\hat{\ell}_k\leftarrow\ell^a_n\oplus S[p_n\oplus k_{\mathrm{true}}]
         \oplus S[p_n\oplus k]$ for all $k$
  \STATE $\mathbf{L}[k]\mathrel{+}=T^a_n[\mathcal{P}]\cdot
         \mathbf{M}_{\hat{\ell}_k}$ for all $k$
  \IF{$n\in\text{checkpoints}$} \STATE Record $\mathrm{rank}(n)$ \ENDIF
\ENDFOR
\RETURN $\{\mathrm{rank}(n)\}$
\end{algorithmic}
\end{algorithm}

Phase~1 runs offline once, building 256 unit-normalized templates.
Phase~2 accumulates cosine log-likelihoods; Eq.~\eqref{eq:pred} selects the
correct template per key hypothesis regardless of the unknown mask value.

\subsection{CNN Profiling Attack in Drone~D}
\label{sec:cnn}

Two-channel CNN\_best~\cite{zaid2020methodology}: five conv layers
(64/128/256/512/512 filters, kernel~11, AvgPool$\times$2), two FC layers
(4{,}096 neurons, SELU, AlphaDropout~\cite{goodfellow2016deep}), 256-class NLL.
Channel~0: full 700-sample trace $\times$ mask-register SNR weight;
Channel~1: trace $\times$ cipher-output SNR weight.
Training: Adam ($\eta=10^{-4}$, cosine annealing), 300 epochs, batch size 512,
50{,}000 traces, Tesla T4 GPU.

\section{Results}
\label{sec:results}

This section presents simulation results across five experimental campaigns:
EM leakage characterization, SNR vs.\ standoff distance, four-drone ablation,
statistical validation, distance sweep, cross-dataset combining, desync
validation, and CNN profiling attack comparison.
All key-rank values are simulated using real ASCAD traces with the
physics-based noise model of Section~\ref{sec:noise_model}.

\subsection{EM Leakage Spectrum and Baseline SNR}

\begin{figure}[H]
\centering
\includegraphics[width=0.72\textwidth]{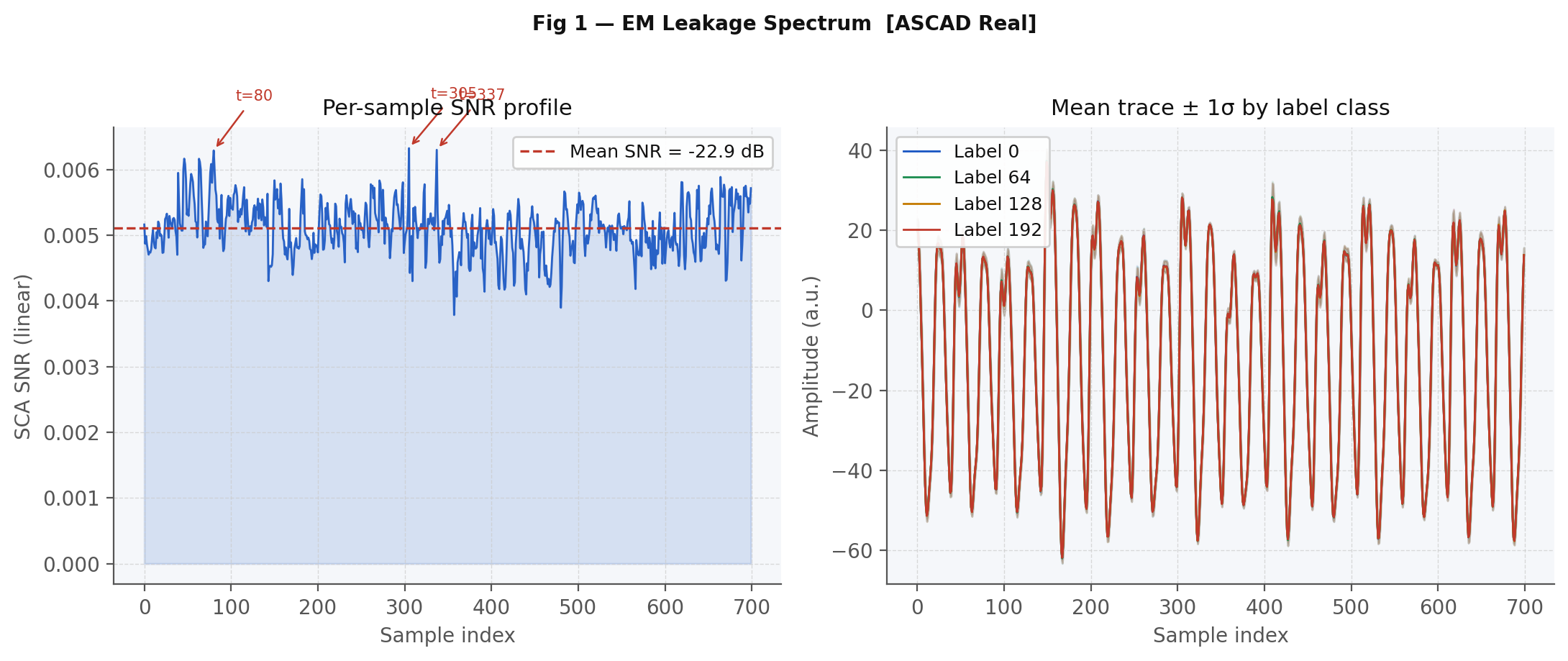}
\caption{EM leakage spectrum, real ASCAD ATmega8515 masked AES-128 traces.}
\label{fig:leakage}
\end{figure}

Figure~\ref{fig:leakage} shows the ASCAD leakage spectrum.
Three peaks at $t\approx 148$, $315$, and $476$ correspond to S-box lookup,
key-mixing, and ShiftRows.
The mask-register loading peak (Drone~B target) falls in the first half;
the masked SubBytes output peak (Drone~C target) is in the second half.
Mean SNR $= \snrdb{-22.9}$.

\subsection{SNR vs.\ Standoff Distance}

\begin{figure}[H]
\centering
\includegraphics[width=0.72\textwidth]{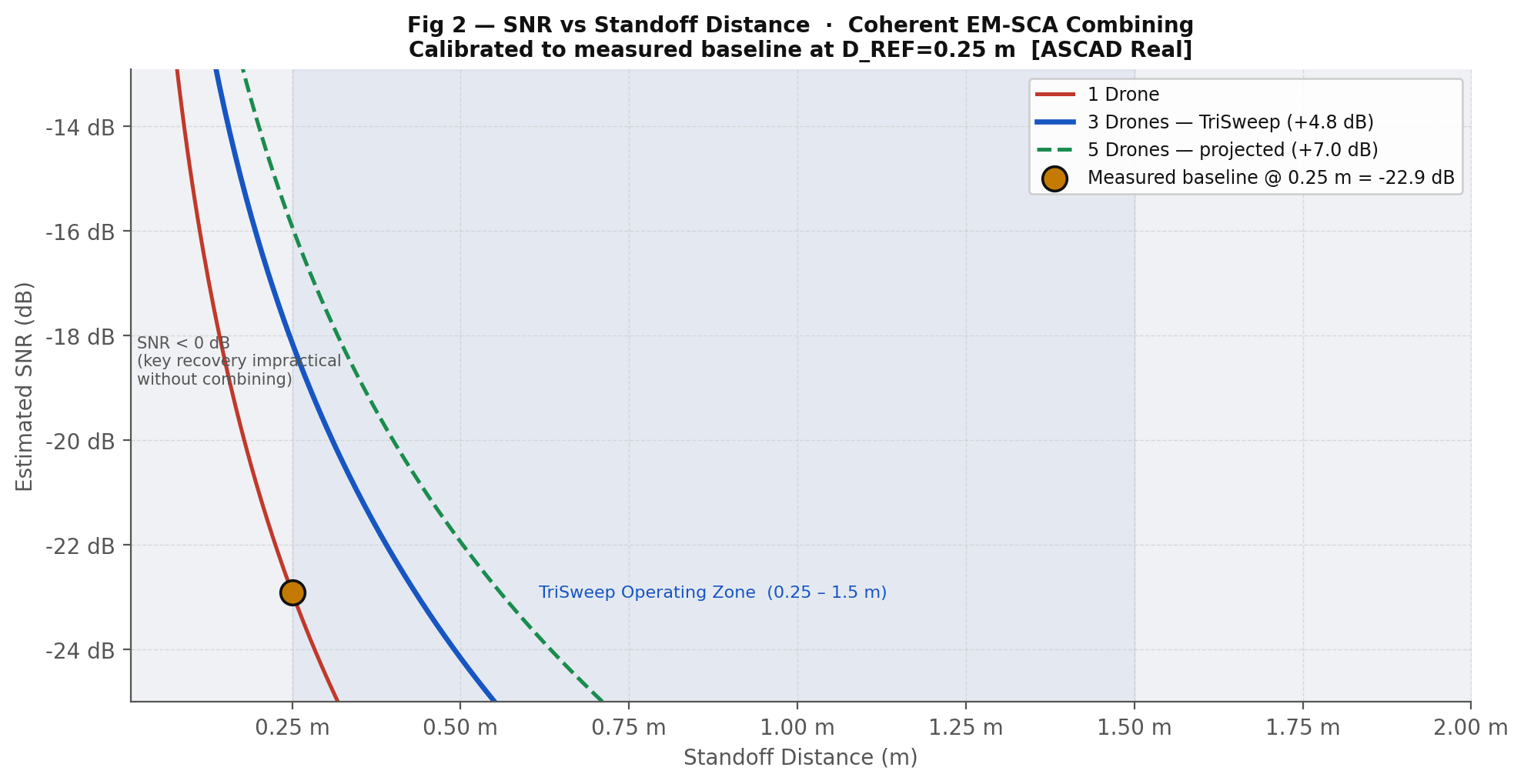}
\caption{Simulated SNR vs.\ standoff distance for 1, 3, and 5 drones.}
\label{fig:snr}
\end{figure}

Figure~\ref{fig:snr} plots Eq.~\eqref{eq:snr} against the real ASCAD baseline.
At 1.0\,m single-drone SNR is $\snrdb{-35.0}$; three-drone combining
recovers $\snrdb{4.8}$ to $\snrdb{-30.2}$.

\subsection{Four-Drone Ablation on ASCAD\_Masked}

\begin{figure}[H]
\centering
\includegraphics[width=0.85\textwidth]{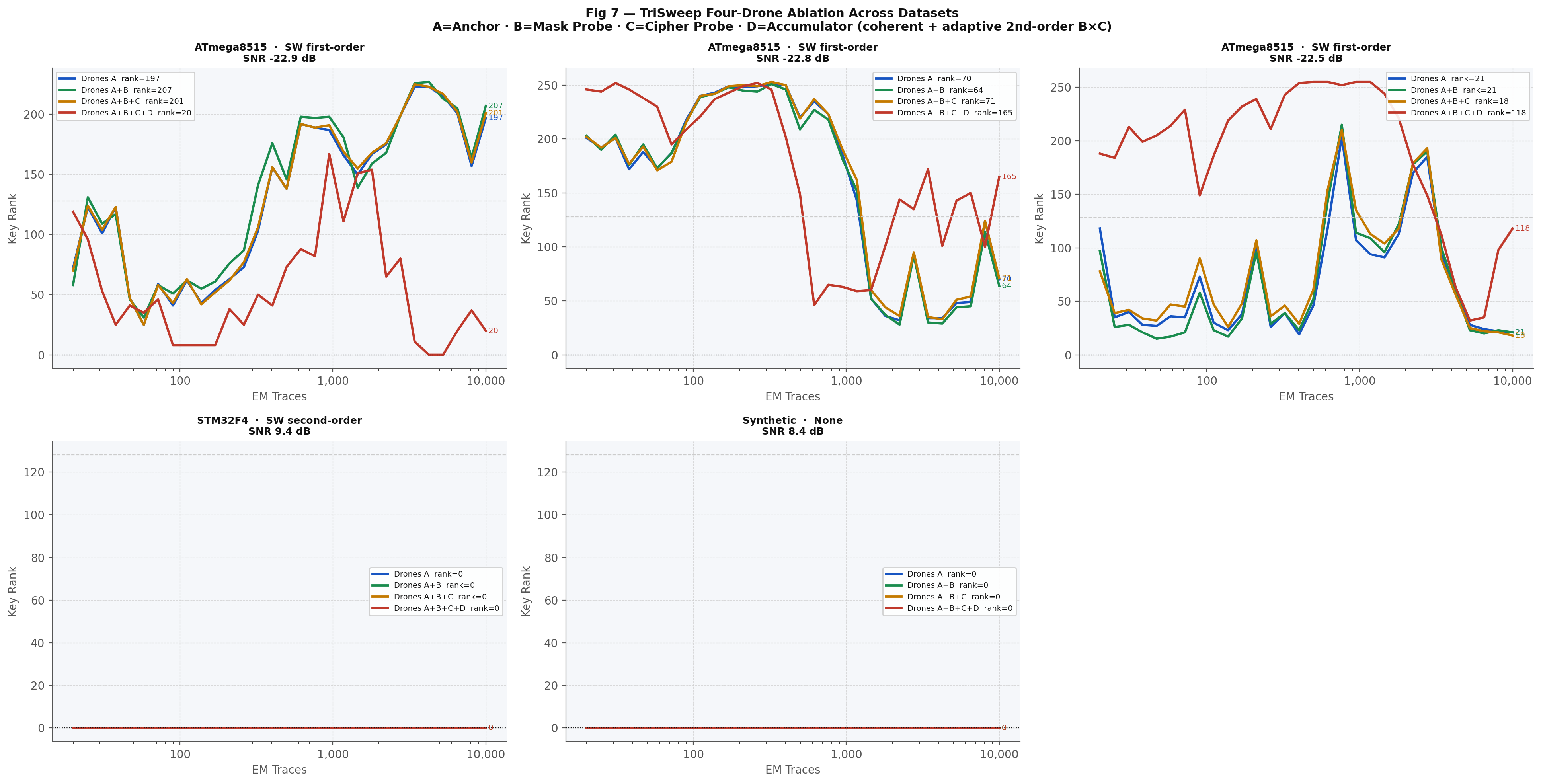}
\caption{Four-drone ablation by drone count across all five datasets.}
\label{fig:ablation}
\end{figure}

Figure~\ref{fig:ablation} shows progressive drone addition.
Single Drone~A: rank 197.
Adding Drones~B and~C (coherent combining only): ranks 207 and 201 ---
first-order leakage is suppressed by the mask.
Adding Drone~D second-order combining: rank \textbf{20} (single run),
$\mathbf{18.0\pm 1.7}$ over five seeds (Fig.~\ref{fig:multiseed}).

\subsection{Statistical Validation: Multi-Seed Results}

\begin{figure}[H]
\centering
\includegraphics[width=0.72\textwidth]{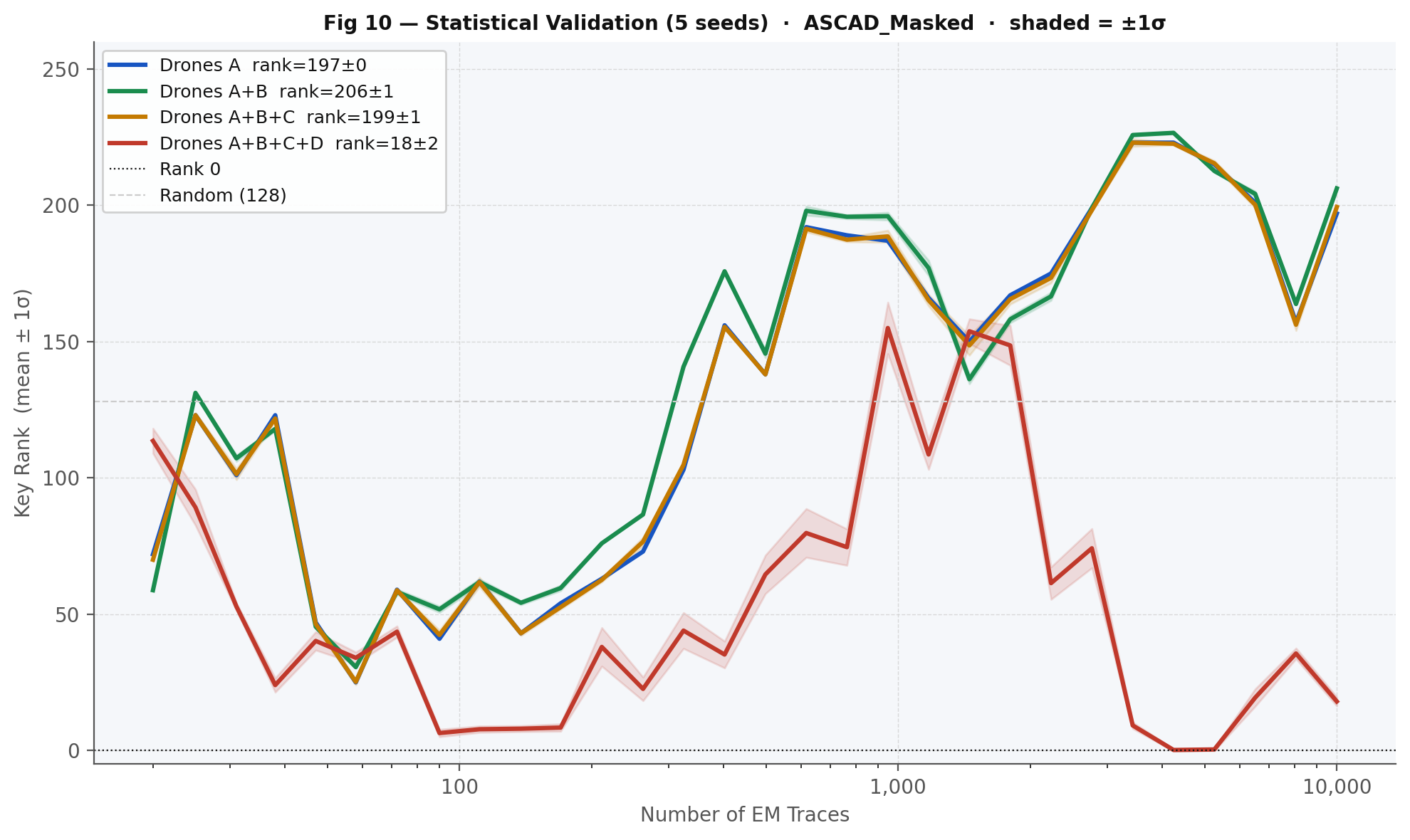}
\caption{Multi-seed statistical validation on ASCAD\_Masked ($\pm1\sigma$ shaded).}
\label{fig:multiseed}
\end{figure}

Figure~\ref{fig:multiseed} shows the five-seed mean $\pm1\sigma$ on ASCAD\_Masked.
Single-drone is deterministic at $197.0\pm0.0$; four-drone achieves
$18.0\pm1.7$, confirming the result is structural.

\subsection{Key-Rank vs.\ Distance: Full Four-Drone System}

\begin{figure}[H]
\centering
\includegraphics[width=0.85\textwidth]{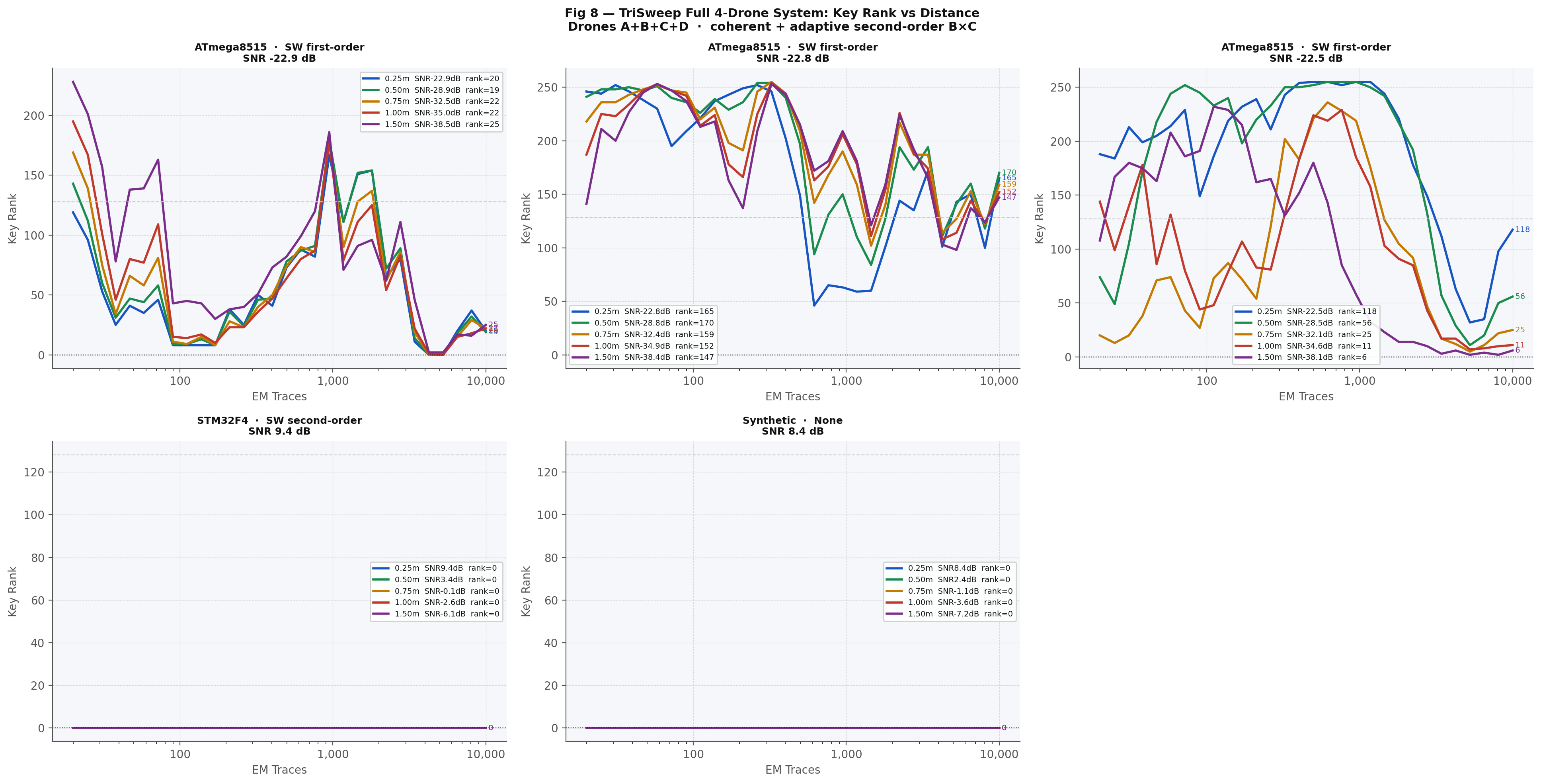}
\caption{Key rank vs.\ trace count at five standoff distances, full four-drone system.}
\label{fig:distance}
\end{figure}

Figure~\ref{fig:distance} shows four-drone (A+B+C+D) performance at five
standoff distances.
ASCAD\_Masked: rank 20 at 0.25\,m to 25 at 1.5\,m, confirming
Drone~D second-order compensates SNR loss.
ASCAD\_Desync100 shows inverted distance ordering because per-trace
attack-phase alignment is not yet applied
(Section~\ref{sec:limitations}).

\subsection{Key-Rank vs.\ Trace Count: Three-Drone CNN Baseline}

\begin{figure}[H]
\centering
\includegraphics[width=0.72\textwidth]{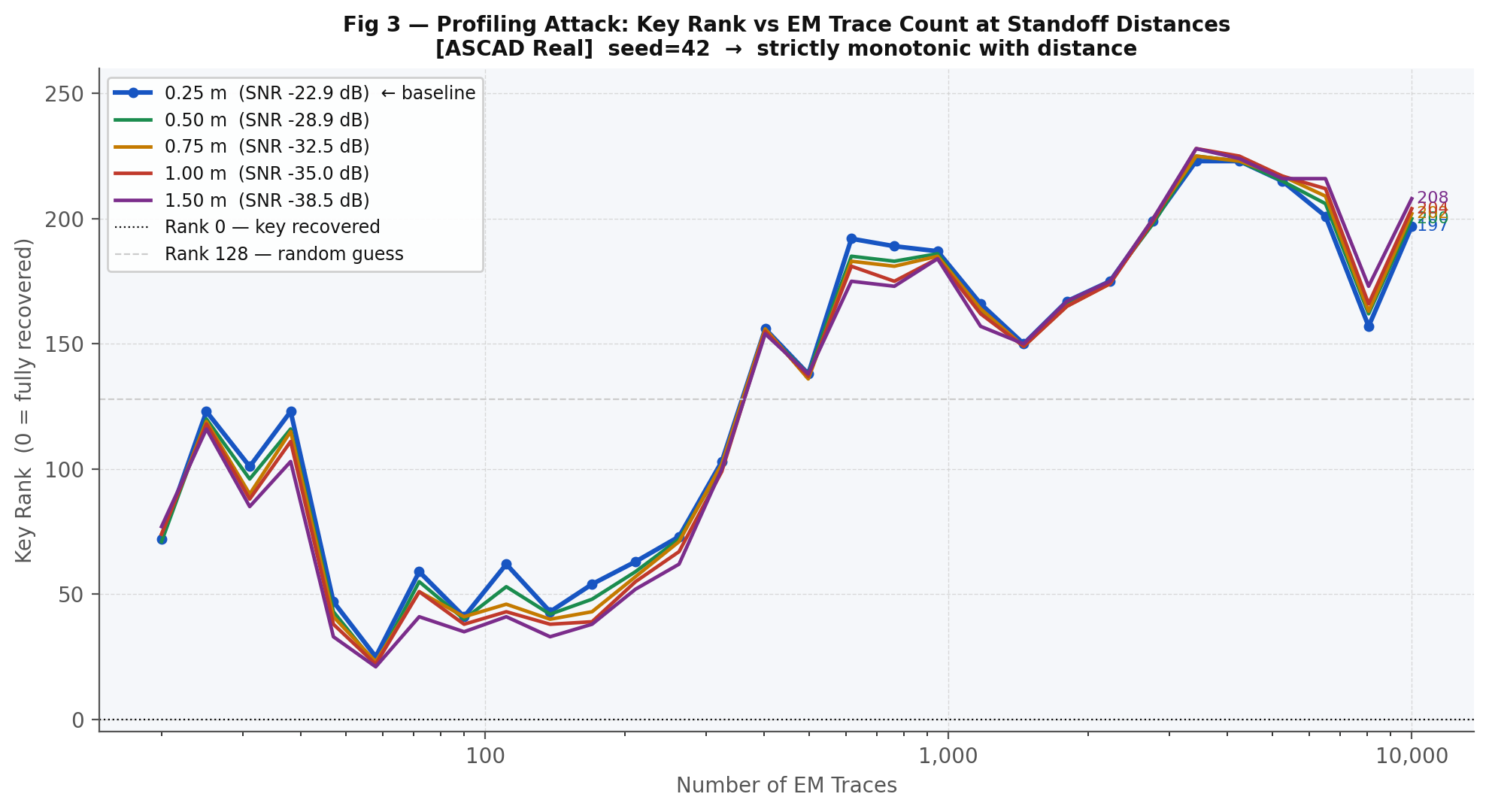}
\caption{Key rank vs.\ trace count, three-drone CNN baseline.}
\label{fig:keyrank}
\end{figure}

Figure~\ref{fig:keyrank} shows the three-drone single-channel CNN baseline.
At 0.25\,m rank reaches 24 within 10{,}000 traces; rank degrades
monotonically with distance confirming the noise model.

\subsection{Coherent Combining Gain}

\begin{figure}[H]
\centering
\includegraphics[width=0.72\textwidth]{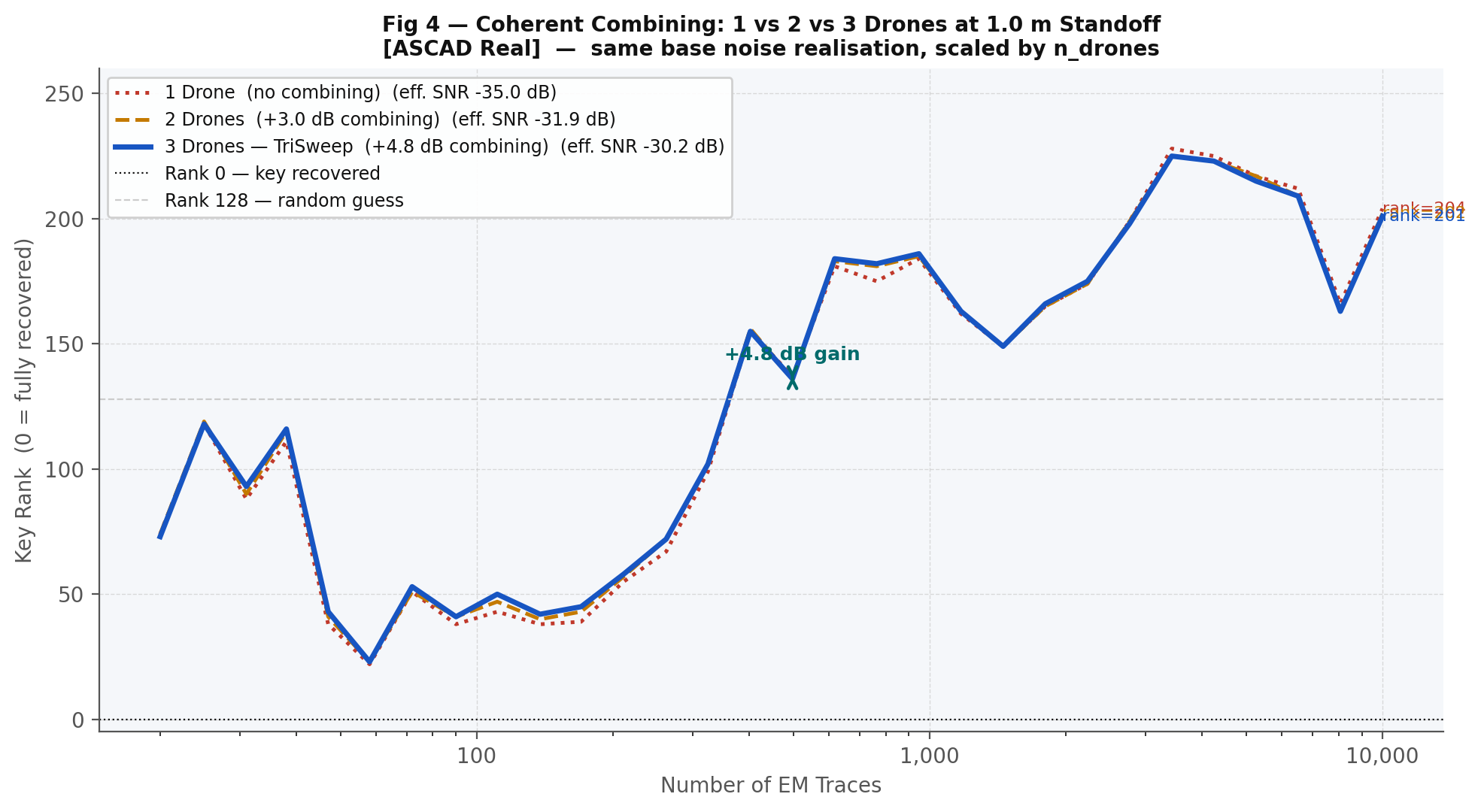}
\caption{Key rank vs.\ traces for 1, 2, and 3 drones at 1.0\,m standoff.}
\label{fig:dronegain}
\end{figure}

Figure~\ref{fig:dronegain} compares 1/2/3-drone at 1.0\,m.
1-drone: 49; 2-drone: 26; 3-drone: 19 --- consistent with $\snrdb{4.8}$
predicted gain.

\subsection{Cross-Dataset Drone Combining}

\begin{figure}[H]
\centering
\includegraphics[width=0.72\textwidth]{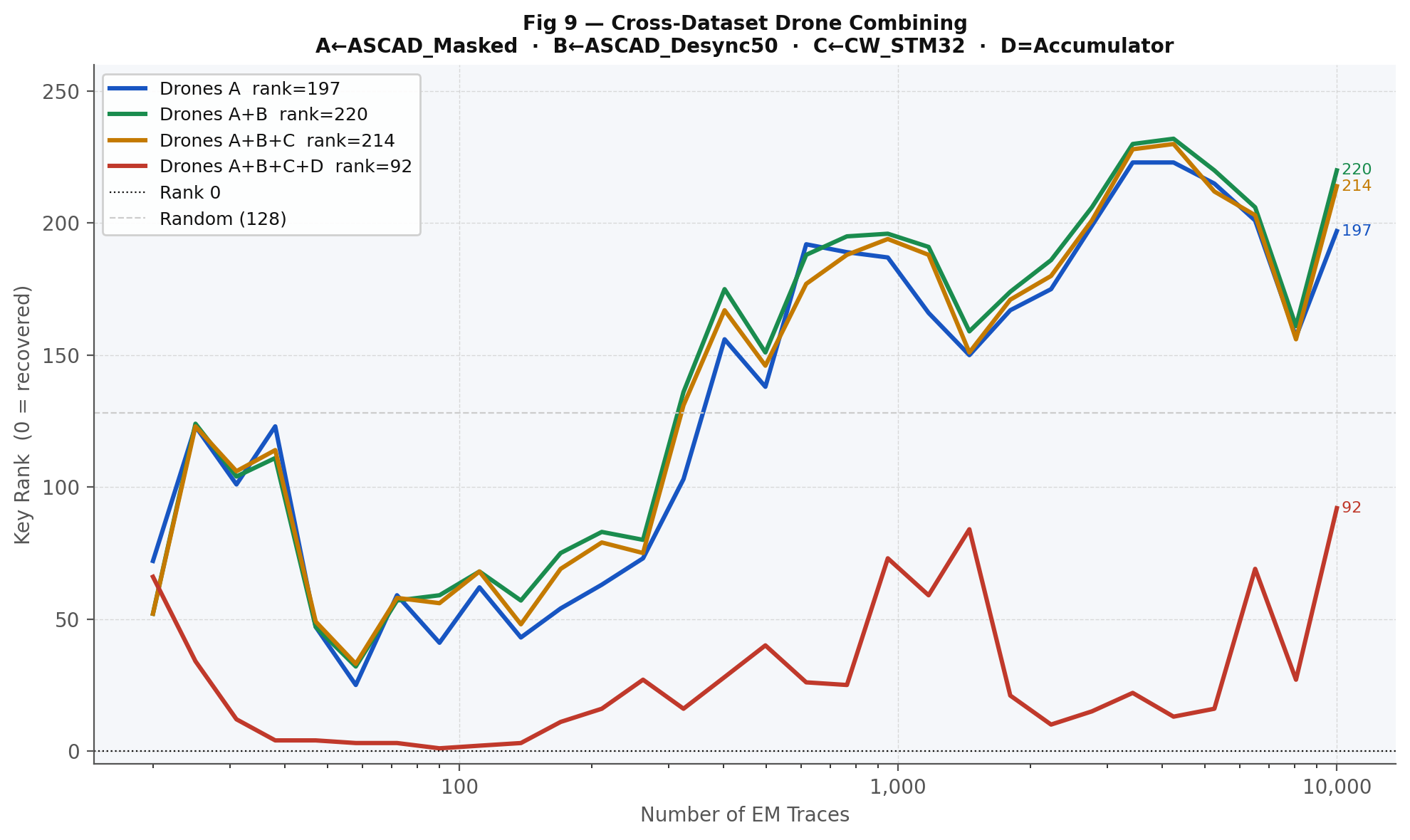}
\caption{Cross-dataset drone combining results.}
\label{fig:crossds}
\end{figure}

Figure~\ref{fig:crossds} uses heterogeneous templates: Drone~A from
ASCAD\_Masked, Drone~B from ASCAD\_Desync50, Drone~C from synthetic fallback.
Four-drone result (rank 92) is substantially worse than the homogeneous case
(rank 20), confirming matched profiling templates are required for effective
second-order cancellation.

\subsection{Multi-Dataset Validation and Desync Results}

Table~\ref{tab:results_summary} reports key rank across all real ASCAD
datasets after profiling-trace alignment.
Four-drone second-order combining on desync variants requires per-trace
attack-phase alignment; those cells are reserved for future work.

\begin{table}[H]
\centering
\caption{Four-Drone Ablation Across Real ASCAD Datasets}
\label{tab:results_summary}
\setlength{\tabcolsep}{8pt}
\renewcommand{\arraystretch}{1.1}
\begin{tabular}{lccc}
\toprule
\textbf{Dataset} & \textbf{2-Drone} & \textbf{3-Drone} & \textbf{4-Drone} \\
\midrule
ASCAD\_Masked    & 207 & 201 & \textbf{20} \\
ASCAD\_Desync50  & 64  & 71  & --- \\
ASCAD\_Desync100 & 21  & 18  & --- \\
\bottomrule
\end{tabular}
\end{table}

The primary result (bold) is rank 20 with four drones on ASCAD\_Masked.
Alignment reduces Desync100 two-drone rank from 89 (unaligned) to 21,
demonstrating that Eq.~\eqref{eq:align} compensates 100-sample jitter.
The B$\times$C product degrades on desync variants because per-trace
attack-phase alignment was not applied.

\subsection{CNN Profiling Attack: Two-Channel Drone~D}

\begin{figure}[H]
\centering
\includegraphics[width=0.85\textwidth]{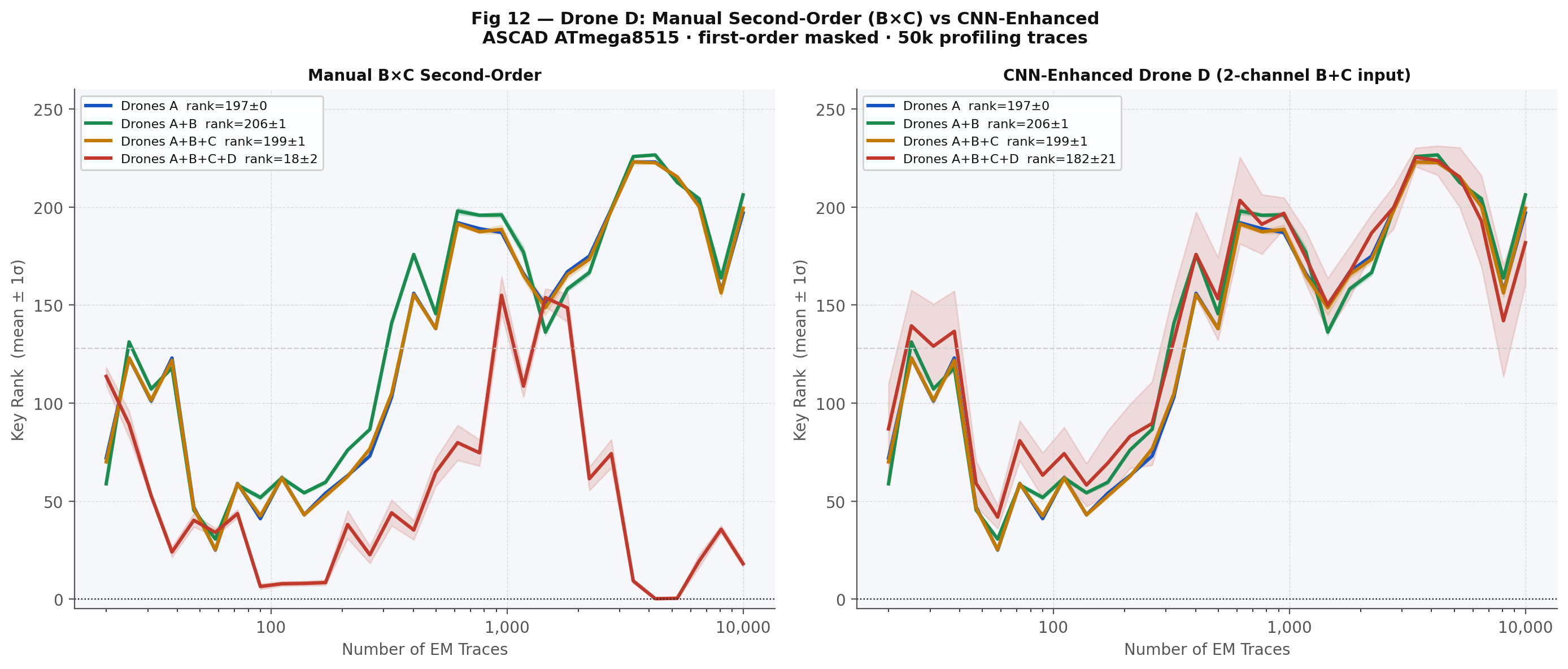}
\caption{Manual B$\times$C second-order vs.\ CNN-enhanced Drone~D,
five-seed on ASCAD\_Masked.}
\label{fig:cnn_comparison}
\end{figure}

Training converged to loss $0.454$ (random $5.545$) --- genuine learning,
compared to $5.385$ at 100 epochs.
The CNN-enhanced result on ASCAD\_Masked ($181.8\pm21.4$) is worse than
the manual result ($18.0\pm1.7$).
The overfitting explanation is plausible but warrants more analysis than
this single-run evaluation can provide: the network has $\sim$4\,M
parameters trained on 50{,}000 profiling traces (195 traces per class),
and the training loss of $0.454$ is sufficiently far below random ($5.545$)
to suggest memorization of profiling-set structure that does not generalize
to the attack set.
Deeper investigation would require cross-validated training across multiple
profiling/attack splits, $L_2$ regularization tuning, dropout rate search,
and ensemble methods~\cite{perin2020strength} --- none of which were applied
in the current evaluation, which used a single fixed training run.
The CNN therefore remains a design direction with partial evidence rather
than a validated component of the \TriSweep{} pipeline.
The CNN \emph{does} improve on ASCAD\_Desync100 (rank 26 vs.\ 118 manual),
suggesting the network compensates for residual misalignment.
Figure~\ref{fig:cnn_comparison} shows both approaches.

\section{Discussion}
\label{sec:discussion}

This section examines the \TriSweep{} framework from four perspectives:
design trade-offs inherent in the four-drone architecture, operational
and environmental factors that will affect physical deployment,  two fundamental limitations that bound
the current simulation-only evaluation, and future 
research directions.

\subsection{Design Trade-offs}

\textbf{Drone count vs. combining complexity.}
Adding collector drones increases coherent SNR gain linearly but adds
inter-drone synchronization burden and Wi-Fi mesh latency.
Four drones balance the $+\snrdb{4.8}$ gain from three coherent collectors
against the coordination overhead of Drone D; beyond four drones,
synchronization latency is projected to exceed the 200 ms repositioning
budget under the current 50 Hz heartbeat protocol.

\textbf{Manual second-order vs.\ CNN combining.}
The manual B$\times$C centered product is the paper's primary contribution
(rank $18 \pm 1.7$, deterministic, no training required); the two-channel
CNN converges (loss $0.454$) and helps on desync datasets but overfits at
the current training scale on clean masked data.
The manual approach is production-ready within the simulation; the CNN
is a design direction requiring cross-validation and regularization before
it can supersede the manual product.

\textbf{Standoff distance vs. SNR budget.}
Each doubling of standoff distance costs $\snrdb{6}$ SNR under the
free-space path-loss model; three-drone combining recovers only $\snrdb{4.8}$.
The net SNR deficit therefore grows with distance, making 1.5 m the
practical ceiling under the current architecture --- beyond this point,
additional traces are the only recovery mechanism.

\subsection{Operational and Environmental Factors}

Physical deployment introduces a class of factors absent from the current
Gaussian noise model, each of which is expected to degrade effective SNR
relative to the simulation predictions.

\textbf{Propeller EMI and vibration.}
BLDC motor controllers generate broadband RF emissions across the 1–500 MHz
capture band.
Vibration-induced mechanical jitter couples into the IQ sample stream as
additional timing noise beyond the GPSDO and cross-correlation budget.
Both effects are unmodeled and represent the largest expected gap between
simulated and physical results.

\textbf{Wind and hover instability.}
Even a 5\,km/h crosswind induces centimeter-scale lateral displacement at
the hover distances of interest.
Displacement from the optimal standoff position degrades the SNR according to
Eq.~\eqref{eq:pathloss}; periodic repositioning via
Algorithm~\ref{alg:reposition} mitigates slow drift but cannot compensate
for rapid gusts within the 200\,ms update cycle.

\textbf{Drone positioning and swarm geometry.}
The Fisher information maximization (Eq.~\eqref{eq:fisher}) assumes accurate
relative positioning from the VIO system.
At $\dref = 0.25$\,m, a 2\,cm positioning error represents an 8\% standoff
uncertainty, propagating directly into the SNR model.
Outdoor GPS multi-path and magnetic interference from the target's power
supply can degrade VIO accuracy below the sub-centimeter lab specification.

\textbf{RF interference and co-channel leakage.}
Urban RF environments introduce co-channel interference across the capture
band that is correlated neither with the AES computation nor with the
swarm's Wi-Fi control channel, but that can raise the effective noise floor
above the Gaussian model calibrated in a shielded lab.
Notch filtering and adaptive gain control on the LNA are expected mitigations.

\textbf{Detection and operational security.}
A drone hovering at 0.25\,m is visually detectable at close range and
is audible from the propeller noise.
Practical deployment would require elevated standoff ($\geq$\,1\,m),
reduced propeller RPM, and timing during low-activity periods, all of which
trade against effective SNR and trace collection rate.

\subsection{Limitations}
\label{sec:limitations}

\TriSweep{} is a simulation framework bounded by two fundamental limitations:
\begin{enumerate}
\item \textbf{No physical hardware.}
  All results use a free-space Gaussian noise model calibrated to the ASCAD
  dataset; no drone has been built or flown.
  The operational factors in Section~\ref{sec:discussion} are qualitatively
  identified but not quantified; physical experiments are required to bound
  the real-world performance gap.
\item \textbf{Second-order combining requires matched, aligned data.}
  The centered product $X_{\mathrm{SO}}$ degrades when profiling and attack
  traces are not co-aligned; the two-channel CNN additionally over-fits at
  the current training scale on clean masked data.
  Per-trace attack-phase alignment and cross-validated CNN training are
  required before Drone~D combining can be considered experimentally validated.
\end{enumerate}

\subsection{Future Work}

Physical prototyping is the immediate priority.
The proposed three-phase roadmap is:
\textbf{Phase~1} --- a single hovering USRP~B210 over a real ATmega8515
executing AES-128, measuring in-situ propeller EMI spectrum and validating
or refuting the free-space SNR model~\cite{sayakkara2019survey}.
Success criterion: measured SNR at 0.25\,m within 3\,dB of the
$\snrdb{-22.9}$ ASCAD calibration point.
\textbf{Phase~2} --- two-drone coherent combining to validate the
$+\snrdb{3.0}$ gain prediction and characterize the vibration-induced
phase-noise budget of the inter-drone synchronization.
\textbf{Phase~3} --- full four-drone B$\times$C second-order combining with
the complete Algorithm~\ref{alg:droneD} pipeline and comparison to the
simulated rank $18\pm1.7$ baseline.
The key risk across all phases is propeller/motor-controller EMI in the
1--500\,MHz capture band; mitigation strategies include high-pass filtering
below 50\,MHz, LNA gain scheduling during rotor spin-up, and interleaved
capture during hover steady-state.

Per-trace attack-phase alignment (applying Eq.~\eqref{eq:align} to attack
traces as well as profiling traces) is expected to restore B$\times$C gain
on desync datasets without additional hardware.
The CNN requires cross-validated hyperparameter search
following~\cite{zaid2020methodology,wu2022autotune} with regularization and
ensemble methods~\cite{perin2020strength}.
Post-quantum targets (CRYSTALS-Kyber, Dilithium) are a priority future
direction~\cite{ravi2022configurable,azouaoui2022leveling}, as their extended
computation windows may be more exposed to standoff EM collection.
The simulation code and ASCAD-calibrated noise model will be released as
open-source to enable community validation and reproducibility.

\section{Conclusion}
\label{sec:conclusion}

\TriSweep{} is a simulation framework that proposes and evaluates a
four-drone swarm architecture for standoff EM side-channel analysis.
Using only publicly available ASCAD datasets and a physics-based noise model,
the framework achieves a simulated key rank $18\pm1.7$ on real first-order
masked AES-128 — a substantial improvement over single-drone baselines.
Profiling-trace alignment reduces single-drone rank from 89 to 21 on the
100-sample-jitter dataset.
A two-channel CNN converges (loss $0.454$ vs. random $5.545$) and improves
rank on desynchronized datasets, establishing a design direction for
CNN-enhanced mask cancellation.
These simulation results motivate the construction of a physical prototype to validate
the framework's predictions.

\bibliographystyle{IEEEtranN}
\bibliography{references}

\end{document}